\documentclass[cameraready]{Interspeech}

\usepackage{cite}

\title{Who is Speaking or Who is Depressed? \\ A Controlled Study of Speaker Leakage in Speech-Based Depression Detection}

\author[affiliation={1},orcid=0009-0004-5613-4814]{Hsiang-Chen}{Yeh}
\author[affiliation={2}, orcid=0009-0004-4702-0508]{Luqi}{Sun}
\author[affiliation={2},orcid=0009-0001-3057-2021]{Aurosweta}{Mahapatra}
\author[affiliation={2}, orcid=0000-0002-1085-2544]{Shreeram Suresh}{Chandra}
\author[affiliation={3}, orcid=0000-0003-1870-6063]{Emily~Mower~Provost}{}
\author[affiliation={2}, orcid=0000-0001-8078-3305]{Berrak~Sisman}{}

\address{
    $^1$ Clinical Mental Health Counseling, Johns Hopkins University, USA \\
    $^2$ Center for Language and Speech Processing, Johns Hopkins University, USA \\
    $^3$ University of Michigan, Ann Arbor, Michigan, USA
}

\email{hyeh10@jh.edu, sisman@jhu.edu}


\usepackage{comment}
\usepackage{subcaption}
\usepackage{soul}
\usepackage{xcolor}
\usepackage{multirow}

\begin{document}

\maketitle

\begin{abstract}

This study investigates whether speech-based depression detection models learn depression-related acoustic biomarkers or instead rely on speaker identity cues. Using the DAIC-WOZ dataset, we propose a data-splitting strategy that controls speaker overlap between training and test sets while keeping the training size constant, and evaluate three models of varying complexity. Results show that speaker overlap significantly boosts performance, whereas accuracy drops sharply on unseen speakers. Even with a Domain-Adversarial Neural Network, a substantial performance gap remains. These findings indicate that depression-related features extracted by current speech models are highly entangled with speaker identity. Conventional evaluation protocols may therefore overestimate generalization and clinical utility, highlighting the need for strictly speaker-independent evaluation.

\end{abstract}


\section{Introduction}

Major Depressive Disorder affects over 332 million people worldwide, with an estimated burden exceeding 56 million disability-adjusted life years ~\cite{herrman2022time, rong2025global}. Despite this burden, most affected individuals remain undiagnosed or untreated due to systemic healthcare barriers and limited access to mental health professionals~\cite{corrigan2004stigma, moitra2022global, wang2005failure}. As a result, automated speech-based depression detection has emerged as a promising objective screening tool to support clinical assessment.

Over the past decade, speech-based depression detection systems have progressed rapidly~\cite{leal2024speech, liu2024diagnostic}. Recent deep learning architectures report classification accuracies exceeding 90\% on benchmark datasets such as DAIC-WOZ~\cite{valstar2016avec, gratch2014distress}. Rezaee et al.~\cite{rezaee2026depression} proposed a streamlined ResNet enhanced with Temporal-Frequency-Channel Attention (TFCA) directly on raw speech, achieving 93.9\% accuracy. Gupta et al.~\cite{gupta2024radiance} introduced RADIANCE, a Transformer-based architecture that captures deep temporal dependencies in speech features. Huang et al.~\cite{huang2024depression} fine-tuned Wav2Vec 2.0~\cite{baevski2020wav2vec} with a customized classification network, reaching 96.5\% accuracy.

Despite these architectural advancements, a critical vulnerability remains. While numerous recent studies report classification accuracies exceeding 90\%, deploying these models on unseen patients often yields performance near random chance~\cite{berisha2024responsible, danylenko2025common}. A primary methodological reason driving these inflated metrics is speaker leakage, where recordings from the same individual appear in both the training and testing sets~\cite{berisha2024responsible, ravi2024enhancing}.

This issue is particularly acute in repeated diagnosis settings, where a patient’s historical recordings are incorporated into training alongside data from other speakers~\cite{berisha2024responsible}. Under such configurations, models face a potential risk of identity confounding when diagnosing patients, as they may inadvertently rely on predictive yet non-causal cues. Rather than solely extracting generalized depression-related acoustic biomarkers, the model might leverage speaker-specific biometric traits to aid classification~\cite{lapuschkin2019unmasking, ravi2024enhancing}. If the network associates a particular voice with a depression label during training, it could reproduce that label at test time by identifying the speaker’s voiceprint. 

To systematically examine this phenomenon, we analyze the behavior of acoustic architectures under controlled clinical training sets. First, we introduce a size-matched data split framework with controlled subject overlap that isolates the effect of speaker overlap while keeping the training size constant, allowing performance differences to be directly attributed to identity leakage. Second, we investigate identity confounding through a progression of three model groups with increasing structural complexity: (i) Wav2Vec-Linear Probing, (ii) XLSR-eGeMAPS Concatenation, and (iii) Wav2Vec-SLS under both frozen and fine-tuned encoder settings, with and without Domain-Adversarial Neural Networks (DANN)~\cite{ganin2016domain}. Through this analytical stress test, we demonstrate that pathological markers and biometric identity are deeply entangled. Across architectures, performance improves dramatically under speaker-overlapped evaluation but drops sharply under strict speaker independence (e.g., 97.65\% to 58.74\% accuracy for a fine-tuned Wav2Vec model). These findings suggest that depression-related signals and speaker identity remain tightly coupled in current speech representations. Our main contributions are summarized as follows:

\begin{itemize}
    \item We propose a data split method with controlled subject overlap that keeps the training set size constant while varying speaker overlap with the test set, enabling a more diagnostic evaluation paradigm for speech-based depression detection.
    \item 
    Through quantitative analysis of speaker identification, we show that the model retains strong identity recognition while achieving high depression classification performance, revealing strong coupling between depression signals and speaker characteristics in current representations.

    \item We systematically evaluate three model architectures of varying complexity, two encoder settings, and multiple configurations with or without DANN, showing that reliance on speaker identity is not model-specific but a common issue in current speech-based depression detection systems.

\end{itemize}


\section{The Proposed Data Split}

\subsection{Dataset and Preprocessing}

We use the Distress Analysis Interview Corpus – Wizard of Oz (DAIC-WOZ) dataset~\cite{valstar2016avec, gratch2014distress}. Each participant completes a single clinical interview lasting 5–20 minutes. Depression severity is assessed using the PHQ-8~\cite{PHQ-8}, with a score $\geq$10 indicating clinical depression. We adopt the standard subset of 189 subjects, comprising 133 healthy controls and 56 depressed participants. Using transcript timestamps, we extract participant-only speech by removing interviewer segments and background silence. Every five consecutive participant utterances are concatenated into a single acoustic segment. This preprocessing yields 6,545 valid speech segments across all 189 subjects.

\vspace{-1mm}
\subsection{Speaker-Overlap Controlled Data Split}
\label{data_split}

\begin{figure}[t!] 
    \centering
    \includegraphics[width=0.9\linewidth]{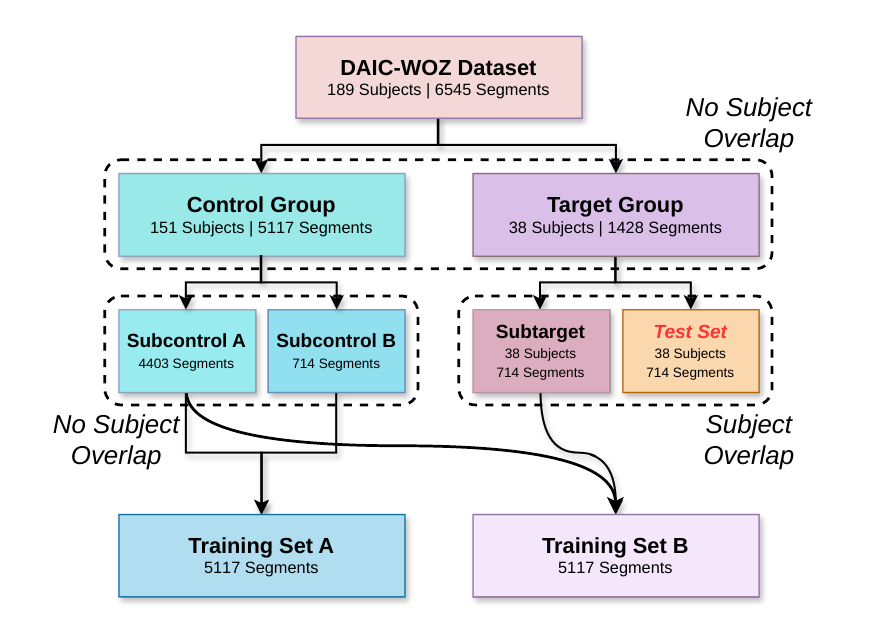}
    \vspace{-4mm}
    \caption{Size-matched data split with controlled subject overlap. (Training Set A: no speaker overlap with test set; Training Set B: speaker overlap with test set.)
    }
    \label{fig:data_split}
    \vspace{-3mm}
\end{figure}

To isolate the effect of speaker overlap while keeping the training scale constant, we design a size-matched data split with controlled subject overlap (Fig.~\ref{fig:data_split}). The 189 participants are first randomly divided into two disjoint speaker groups: a \textit{control group} (151 speakers, 5,117 segments) and a \textit{target group} (38 speakers, 1,428 segments). Here, the term “control group” refers to non-test speakers and should not be confused with healthy control subjects. No speaker appears in both groups. The 1,428 segments from the target group are evenly split within each speaker into two equal subsets (714 segments each). One subset serves as the shared test set for all experiments. The other subset (subtarget) is optionally included in training to simulate speaker overlap. To ensure identical training size across conditions, 4,403 segments are randomly sampled from the control group to form subcontrol A, while the remaining 714 segments form subcontrol B. Two training sets, each containing 5,117 segments, are then constructed:

\begin{itemize}
    \item \textbf{Training Set A (Speaker-Independent Setting):} all 5,117 segments from the control group; no speaker overlap with the test set.
    \item \textbf{Training Set B (Speaker-Overlapped Setting):} subcontrol A (4,403 segments) combined with subtarget (714 segments); speakers overlap with the test set.
\end{itemize}
Thus, the only difference between Training Set A and Training Set B is whether speakers in the test set are partially observed during training, while both the training size and the test set remain fixed.

\vspace{-2mm}
\section{Methodology}

We evaluate three model families of increasing architectural complexity under both speaker-independent (Training Set A) and speaker-overlapped (Training Set B) conditions. Each model is tested in its original form and with a DANN extension to mitigate speaker-specific information. 

\vspace{-1mm}
\subsection{Domain-Adversarial Neural Network}
\label{sec:DANN}
Domain-Adversarial Neural Network (DANN)~\cite{ganin2016domain} is a framework for learning domain-invariant representations by using a Gradient Reversal Layer (GRL). This method improves domain classification error~\cite{ravi2022step, wang2023non} while maintaining the performance of the primary task. In this study, we treat speaker identity as a confounding domain. In the task of speech-based depression detection, factors such as a speaker’s voiceprint characteristics, physiological vocal structure, and expressive habits may form highly distinguishable identity information in the feature space. When there is speaker overlap between the training and test sets, the model may exploit these identity cues as shortcuts for prediction, thereby leading to overestimated performance. To mitigate this potential identity dependence, we incorporate DANN into each model architecture, enabling the encoder to remain insensitive to speaker identity while learning depression-discriminative features. Through this mechanism, we can systematically evaluate a key question: whether the model’s depression detection performance can remain stable after speaker-related features are suppressed.

\begin{figure*}[t!]
\centering
\includegraphics[width=0.85\textwidth]{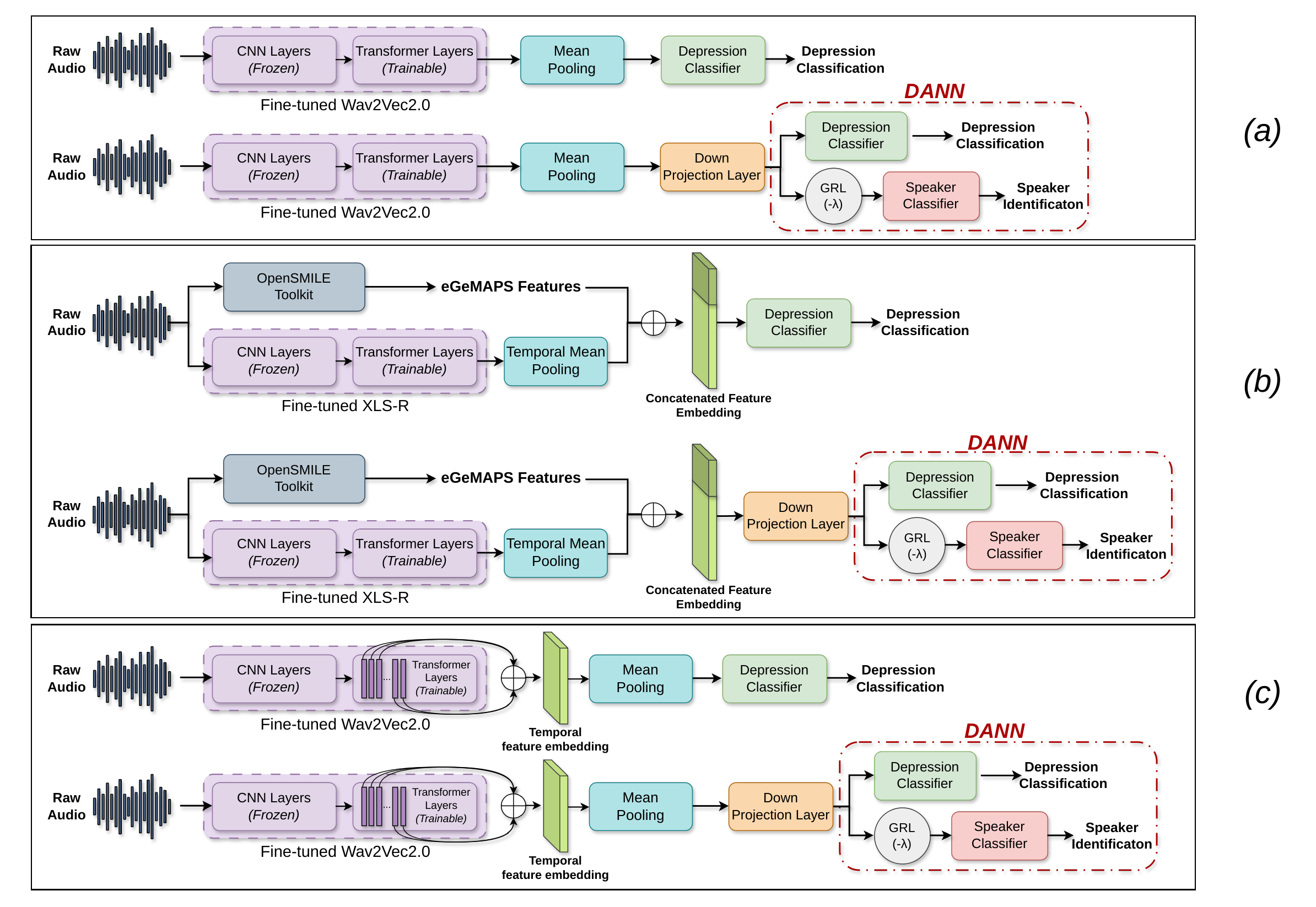}
\caption{Architectures of the three model groups:
(a) Wav2Vec-Linear Probing Models;
(b) XLSR-eGeMAPS Concatenation Models;
(c) Wav2Vec-SLS Models.
(GRL: Gradient Reversal Layer;
DANN: Domain-Adversarial Neural Network.)}
\label{fig:architectures}
\vspace{-3mm}
\end{figure*}

\vspace{-1mm}
\subsection{Wav2Vec-Linear Probing Models}

As shown in Fig.~\ref{fig:architectures}(a), the Wav2Vec-Linear Probing model is built upon Wav2Vec 2.0~\cite{baevski2020wav2vec} to extract speech representations. Raw audio is first processed by the convolutional feature encoder, whose parameters are kept frozen to mitigate catastrophic forgetting~\cite{forget} and reduce overfitting. The resulting representations are then passed through transformer layers, which are fine-tuned to learn task-specific information. Mean pooling is applied to the final hidden representations to obtain a fixed-dimensional embedding~\cite{balagopalan2021comparing}. This embedding is fed into a single-layer linear classifier for depression prediction. To reduce the model’s reliance on speaker identity information, we further extend the original architecture with a DANN-enhanced module. The embedding obtained through mean pooling is first passed through a linear layer for dimensionality reduction, and then fed into the main depression classifier and an adversarial speaker classifier, where the speaker classifier is adversarially optimized through a gradient reversal layer.

\vspace{-1mm}
\subsection{XLSR-eGeMAPS Concatenation Models}
As shown in Fig.~\ref{fig:architectures}(b), the XLSR-eGeMAPS Concatenation model combines self-supervised speech representations with hand-crafted acoustic features. XLS-R~\cite{XLSR}, a multilingual extension of Wav2Vec 2.0, is used to extract contextual embeddings. Raw audio is processed by frozen convolutional layers followed by fine-tuned transformer layers, and temporal mean pooling is applied to obtain a fixed-dimensional representation. In parallel, eGeMAPS features~\cite{egemaps} are extracted using the OpenSMILE toolkit~\cite{opensmile}. The pooled XLS-R embedding and the eGeMAPS features are concatenated to form the final feature representation~\cite{balagopalan2021comparing}, which is fed into a single-layer linear classifier for depression prediction. In the DANN-enhanced variant, the concatenated features are first projected through a linear layer for dimensionality reduction. The projected representation is then used for both depression classification and adversarial speaker classification via a gradient reversal layer, encouraging speaker-invariant feature learning.

\subsection{Wav2Vec-SLS Models} 
As illustrated in Fig.~\ref{fig:architectures}(c), the Wav2Vec-SLS model extends Wav2Vec 2.0 with Sensitive Layer Selection (SLS)~\cite{zhang2024audio}. Raw audio is processed by a frozen convolutional encoder followed by fine-tuned transformer layers. Instead of using only the final layer representation, this model extracts representations from all transformer layers and aggregates them through a weighted summation to capture multi-level acoustic and semantic information. Mean pooling is then applied to obtain a fixed-dimensional representation. The pooled representation is fed into a single-layer linear classifier for depression prediction. In the DANN-enhanced variant, the pooled representation is first projected through a linear layer for dimensionality reduction. The projected features are then used for both depression classification and adversarial speaker classification via a gradient reversal layer, encouraging speaker-invariant feature learning.

\vspace{-1mm}
\section{Experiments}

We evaluate two clinical scenarios: \textit{initial diagnosis} and \textit{repeated diagnosis}. To simulate these under controlled conditions, we use the size-matched split described in Section~\ref{data_split}, yielding two training sets and a shared test set. In the initial diagnosis scenario (Training Set A), the training set contains 5,117 segments with no speaker overlap with the test set. In the repeated diagnosis scenario (Training Set B), the training set also contains 5,117 segments but includes partial data from speakers in the test set. The shared test set consists of 714 segments from 38 speakers in the Target Group. As the training size and test data are identical across conditions, any performance difference is attributable solely to the presence or absence of speaker overlap. We report three metrics: (1) Depression Macro F1-Score~\cite{f1}, (2) Depression Classification Accuracy~\cite{depaccuracy}, and (3) Speaker Identification Accuracy~\cite{spkacc}.

\begin{table*}
\centering 
\caption{Performance under speaker-independent (Training Set A) and speaker-overlapped (Training Set B) settings. Dep Macro F1: Depression Macro F1-Score; Dep Cls Acc: Depression Classification Accuracy; Spk ID Acc: Speaker Identification Accuracy. }
\label{tab:results} 
\resizebox{\linewidth}{!}{
\begin{tabular}{l c l l c c c} 
\toprule


 \textbf{Model Architecture} & \textbf{Encoder} & \textbf{Variant} & \textbf{Training Set} & \textbf{Dep Macro F1 $\uparrow$} & \textbf{Dep Cls Acc $\uparrow$} & \textbf{Spk ID Acc $\downarrow$} \\
\midrule 

\multirow{8}{*}{\shortstack{\textit{\textbf{Wav2Vec-Linear}} \\ \textit{\textbf{Probing Models}}}}

& \multirow{4}{*}{\centering \shortstack{\textit{Frozen} \\ \textit{Wav2Vec 2.0}}}
& \multirow{2}{*}{\centering \textit{Original}}
& A (No Speaker Overlap) & 0.5277 & 54.06\% & 0.00\% \\
& & & B (Speaker Overlap) & 0.7646 & 76.75\% & 95.94\% \\
\cmidrule(lr){3-7}
& & \multirow{2}{*}{\centering \textit{DANN-Enhanced}}
& A (No Speaker Overlap) & 0.5593 & 57.59\% & 0.00\% \\
& & & B (Speaker Overlap) & 0.7546 & 75.85\% & 93.78\% \\
\cmidrule(lr){2-7}

& \multirow{4}{*}{\centering \shortstack{\textit{Fine-tuned} \\ \textit{Wav2Vec 2.0}}}
& \multirow{2}{*}{\centering \textit{Original}}
& A (No Speaker Overlap) & 0.5624 & 58.74\% & 0.00\% \\
& & & B (Speaker Overlap)  & 0.9763 & 97.65\% & 90.95\% \\
\cmidrule(lr){3-7}
& & \multirow{2}{*}{\centering \textit{DANN-Enhanced}}
& A (No Speaker Overlap) & 0.6022 & 62.36\% & 0.00\% \\
& & & B (Speaker Overlap) & 0.9475 & 94.78\% & 67.25\% \\

\midrule


\multirow{8}{*}{\shortstack{\textit{\textbf{XLSR-eGeMAPS}} \\ \textit{\textbf{Concatenation Models}}}}

& \multirow{4}{*}{\centering \shortstack{\textit{Frozen} \\ \textit{XLS-R}}}
& \multirow{2}{*}{\centering \textit{Original}}
& A (No Speaker Overlap) & 0.7098 & 57.28\% & 0.00\% \\
& & & B (Speaker Overlap) & 0.7312 & 62.32\% & 8.26\% \\
\cmidrule(lr){3-7}
& & \multirow{2}{*}{\centering \textit{DANN-Enhanced}}
& A (No Speaker Overlap) & 0.5379 & 59.38\% & 0.00\% \\
& & & B (Speaker Overlap) & 0.6400 & 67.09\% & 6.16\% \\
\cmidrule(lr){2-7}

& \multirow{4}{*}{\centering \shortstack{\textit{Fine-tuned} \\ \textit{XLS-R}}}
& \multirow{2}{*}{\centering \textit{Original}}
& A (No Speaker Overlap) & 0.5439 & 58.68\% & 0.00\% \\
& & & B (Speaker Overlap) & 0.6426 & 66.99\% & 4.62\% \\
\cmidrule(lr){3-7}
& & \multirow{2}{*}{\centering \textit{DANN-Enhanced}}
& A (No Speaker Overlap) & 0.7077 & 54.76\% & 0.00\% \\
& & & B (Speaker Overlap) & 0.7077 & 54.76\% & 10.36\% \\

\midrule

\multirow{8}{*}{\textit{\textbf{Wav2Vec-SLS Models}}}

& \multirow{4}{*}{\centering \shortstack{\textit{Frozen} \\ \textit{Wav2Vec 2.0}}}
& \multirow{2}{*}{\centering \textit{Original}}
& A (No Speaker Overlap) & 0.6371 & 64.47\% & 0.00\% \\
& & & B (Speaker Overlap) & 0.7565 & 76.26\% & 96.22\% \\
\cmidrule(lr){3-7}
& & \multirow{2}{*}{\centering \textit{DANN-Enhanced}}
& A (No Speaker Overlap) & 0.4591 & 55.90\% & 0.00\% \\
& & & B (Speaker Overlap) & 0.8133 & 79.55\% & 89.36\% \\
\cmidrule(lr){2-7}

& \multirow{4}{*}{\centering \shortstack{\textit{Fine-tuned} \\ \textit{Wav2Vec 2.0}}}
& \multirow{2}{*}{\centering \textit{Original}}
& A (No Speaker Overlap) & 0.7383 & 70.31\% & 0.00\% \\
& & & B (Speaker Overlap) & 0.9830 & 98.31\% & 94.96\% \\
\cmidrule(lr){3-7}
& & \multirow{2}{*}{\centering \textit{DANN-Enhanced}}
& A (No Speaker Overlap) & 0.6593 & 66.57\% & 0.00\% \\
& & & B (Speaker Overlap) & 0.9646 & 96.49\% & 88.66\% \\

\bottomrule
\end{tabular} 
} 
\vspace{-4mm} 
\end{table*}

\vspace{-1mm}
\section{Results and Analysis}

Table~\ref{tab:results} illustrates the performance across the three evaluated architectures in multiple settings. 

\textbf{Performance under the speaker-overlapped setting.}
Under the speaker-overlapped setting (Training Set B), most architectures achieve very high depression classification performance. In the Wav2Vec-Linear Probing model, the Original variant with a frozen encoder attains 76.75\% accuracy, which increases to 97.65\% after encoder fine-tuning. Even with adversarial training, the Fine-tuned DANN model maintains strong performance (Dep Macro F1 = 0.9475; Accuracy = 94.78\%). A similar pattern is observed for the Wav2Vec-SLS architecture, where fine-tuning yields classification accuracy close to 98\%. In contrast, the XLSR-eGeMAPS Concatenation model achieves comparatively lower performance (62\% to 67\% accuracy), although it still improves under the speaker-overlapped condition. Overall, when speakers overlap between training and test sets, high classification accuracy is consistently observed across architectures with sufficient modeling capacity.

\textbf{Performance under the speaker-independent setting.} Under strict speaker-independent evaluation (Training Set A), performance drops consistently and substantially. For example, the Wav2Vec-Linear Probing Fine-tuned Original model’s accuracy decreases from 97.65\% to 58.74\%, and DANN provides only limited recovery. A comparable decline is observed in Wav2Vec-SLS, where depression recognition also deteriorates markedly in the absence of speaker overlap. This degradation is systematic across architectures and training strategies, demonstrating limited generalization to unseen speakers. Models that achieve near-ceiling accuracy under speaker overlap fail to sustain comparable performance when identity cues are unavailable. In contrast, the XLSR-eGeMAPS model shows a smaller gap, but its overall depression accuracy remains moderate (54\% to 59\%). Notably, exceptionally high depression performance occurs only in models that also exhibit strong speaker discrimination, reinforcing the association between identity retention and elevated depression accuracy.

\textbf{The role of speaker identity information in model predictions.} To assess whether identity drives the performance gap, we examine Speaker Identification Accuracy under the speaker-overlapped setting (Training Set B), where random chance is $2.63\%$ ($1/N_s$, $N_s=38$). The contrast is clear. The XLSR-eGeMAPS model achieves only 6\% to 10\% speaker identification accuracy (near chance) and its depression accuracy remains moderate (62\% to 67\%). In contrast, the Wav2Vec-Linear Probing Fine-tuned Original model achieves 90.95\% speaker identification accuracy together with 97.65\% depression accuracy. With DANN, speaker identification decreases to 67.25\%, yet depression accuracy remains high at 94.78\%. This pattern is consistent across model variants: stronger speaker discrimination accompanies higher depression accuracy, whereas near-chance identity recognition is associated with markedly lower performance. The evidence suggests that pretrained speech representations preserve identity-related information, which models leverage under speaker-overlapping splits, thereby inflating depression estimates.

\textbf{Generality and implications.}
Comparisons across three model groups of varying complexity reveal a consistent gap between speaker-overlapping and speaker-independent evaluations, with the former consistently yielding substantially higher performance. This pattern holds across architectures, indicating that identity reliance reflects a property of current speech representations rather than a model-specific limitation, and neither increased structural complexity nor larger parameter counts mitigate this effect. The XLSR-eGeMAPS results further clarify the relationship: when speaker identification approaches chance level, depression accuracy remains moderate (62\% to 67\%), whereas substantially higher depression performance appears only when identity information is preserved. Removing segment-level overlap alone does not prevent identity confounding, and strict speaker-independent partitioning is necessary for performance estimates to reflect depression-related signal rather than speaker memorization, supporting a defensible evaluation of clinical validity.

\vspace{-2mm}
\section{Conclusion}
\vspace{-1mm}

We introduced a size-controlled data split and conducted systematic evaluations under speaker-overlapped and speaker-independent conditions to examine identity leakage in speech-based depression detection. Across architectures and training strategies, the high accuracy observed under speaker overlap decreases substantially when evaluated on unseen speakers, indicating a strong dependence on identity-related cues. The results demonstrate that performance gains under overlapping splits do not necessarily reflect improved modeling of depressive pathology. Therefore, we suggest that, in order to more accurately evaluate the model's capabilities and clinical application value, future studies should not only adopt segment-level splits but also conduct evaluation using rigorous speaker-independent settings.

\section{Acknowledgments}
We thank the Johns Hopkins University Data Science and AI (DSAI) Institute for supporting this research through a faculty startup package.

\section{Generative AI Use Disclosure}
Generative AI tools were employed solely for language polishing of text written by the authors. These tools were not used to generate scientific content, results, experimental designs, analyses, or conclusions. All authors are responsible for the full content of this paper and consent to its submission.

\bibliographystyle{IEEEtran}
\bibliography{mybib}

\end{document}